**Title:**

Mechanism-Resolved PFM of Ferroionic and Ferroelectric Responses in Thickness-Gradient $Hf_{0.5}Zr_{0.5}O_2$ Libraries

**Authors:**

Yu Liu[1*], Yi-Xiu Chen[2], Haotong Liang[2], Ichiro Takeuchi[2], and Sergei V. Kalinin[1,3*]

[1] Department of Materials Science and Engineering, University of Tennessee, Knoxville, Tennessee, 37996, USA

[2] Department of Materials Science and Engineering, University of Maryland, College Park, Maryland 20742, USA

[3] Physical Sciences Division, Pacific Northwest National Laboratory, Richland, Washington, 99354, USA

**Abstract:**

Resolving growth mechanisms and thickness evolution of functional properties is one of the key tasks in materials discovery and optimization involving thin-film materials, traditionally requiring significant experimental budgets. Here we introduce the combination of thickness-gradient libraries and automated scanning probe microscopy as a systematic pathway to elucidate growth modes and disentangle ferroelectric and electrochemical contributions in ferroelectric thin films. As a model system, we explore the $Hf_{0.5}Zr_{0.5}O_2$ (HZO) gradient thin films grown on $La_xSr_{1-x}MnO_3$ (LSMO) bottom electrode thin films. Automated piezoresponse force microscopy, spectroscopy, and lithography reveals that irreversible topographic deformation arises from electrochemical activity at the LSMO surface, whereas reversible phase inversion in HZO reflects ferroelectric switching. Automated topography height-map scans are used to further quantify nucleation density, particle-size evolution, and roughness correlations across the thickness-gradient, demonstrating that improved plume stabilization during growth suppresses interfacial reactions and promotes dense, fine-grained HZO conducive to ferroelectric phase formation. This combined materials-engineering and automated-SPM framework establishes a platform for high-throughput, mechanism-resolved characterization of ferroionic and ferroelectric responses in complex oxide films.

I. **Introduction**

Combinatorial materials libraries enable rapid exploration of multidimensional parameter spaces, including composition[1-6], thickness[7], and growth temperature[8,9], within individual substrates. These libraries allow continuous mapping of structure–property relationships while dramatically reducing experimental cost and variability. For example, temperature-gradient libraries[10-13] have been employed to probe phase formation, crystallization windows, and kinetic pathways during growth and post-deposition processing. As synthesis capabilities develop, however, the central challenge has shifted from library fabrication to reliable, mechanism-resolved characterization. Extracting meaningful functional trends across a gradient requires spatially resolved, quantitative measurements capable of disentangling physical mechanisms.

Piezoresponse force microscopy (PFM) has, over the past two decades, emerged as a uniquely powerful technique for probing polarization dynamics at the nanoscale in ferroelectric materials.[14-16] During its first ten years, research centered almost exclusively on classical ferroelectric systems exhibiting strong electromechanical coupling such as lead zirconate–titanate (PZT),[17-19] strontium-barium niobate and titanate films,[20] barium titanate,[14] nitrides,[21] ferroelectric relaxors,[22-25] and water-soluble ferroelectrics such as guanidinium aluminum sulfate hexahydrate and triglycine sulfate (TGS).[26-28]

With the broader adoption of PFM, however, numerous studies soon reported piezoresponse signals in materials traditionally considered non-ferroelectric such as $SrTiO_3$. The overview of these studies is given by Vasudevan.[29] In parallel, it was recognized as early as 2010 that similar electromechanical contrast can arise purely from field-induced ionic motion under the probe, giving rise to what is now called electrochemical strain microscopy (ESM).[30-32] ESM has since been employed to map ionic dynamics in mixed conductors such as cobaltites,[33,34] solid electrolytes such as ceria,[35,36] and memristive materials such as NiO.[37] Finally, the observations of electromechanical hysteresis loop were reported even for materials such as $LaAlO_3$-$SrTiO_3$,[38] suggesting universality of the electromechanical hysteresis emerging due to intrinsic ferroelectricity, electrochemical polarization, surface ionic motion across broad materials classes. From theory perspective, it has been shown that in proper ferroelectric materials the coupling between bulk ferroelectric responses and surface electrochemistry can yield a continuum of ferroionic states that smear the ferroelectric phase transition, sensitively depending on environmental conditions.[39-48]

It is generally expected that true ferroelectric switching yields well-defined, rectangular hysteresis loops with two possible polarization values while ionic contributions produce slanted or multistep loops with continuum of polarization states. However, ferroionic states and ferroelectric relaxors do not follow this pattern, and a survey of the literature makes it clear that these distinctions are insufficient for unambiguous mechanism assignment without supporting structural or spectroscopic evidence.[29, 48]

In this work, we introduce a materials investigation approach in which combinatorial libraries are fabricated with systematic variations in the film thickness. When coupled with automated PFM, it provides multiple standards for separating genuine polarization switching from electrochemical effects. We demonstrate this approach on hafnia-based epitaxial heterostructures with LSMO electrodes, leveraging automated scanning-probe protocols to enable high-throughput, positionally encoded characterization of electromechanical and electrochemical responses across a single thickness-gradient sample. We show that the electromechanical and electrochemical response can be distinguished through lithography scans, and their competition is controlled by the surface chemistry and thickness of the ferroelectric film. Furthermore, our topography scan revealed the dominance of surface chemistry and thin-film growth mode depending on the deposition condition and the composition of the LSMO electrode, showing the power of thickness-gradient combinatorial libraries in optimizing both film growth parameters and functionalities.

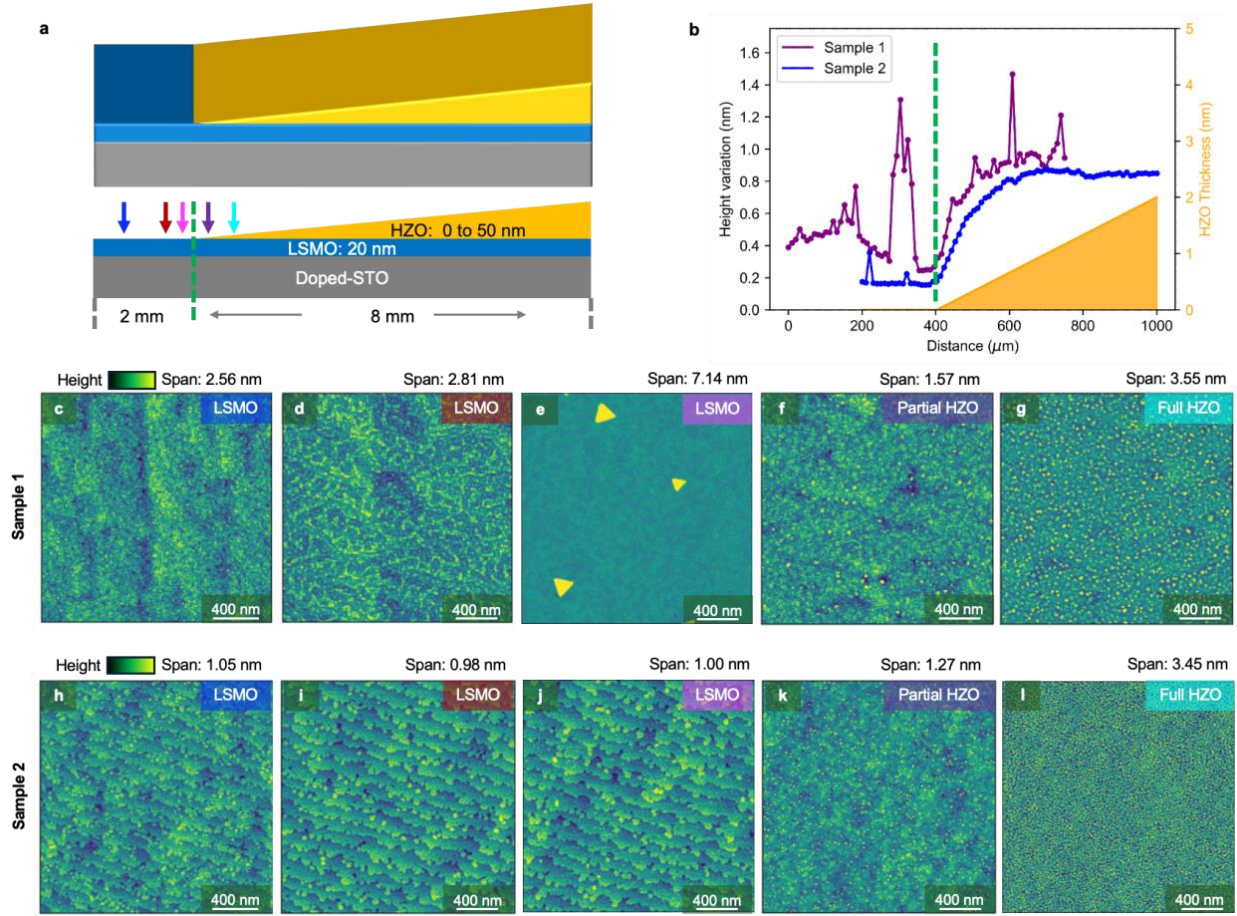

**Figure 1. Overview of the Hf$_{0.5}$Zr$_{0.5}$O$_2$ (HZO) thickness combinatorial library. a,** schematic drawing of the HZO thickness library. In the first 2 mm of the sample, there is exposed LSMO substrate for contrast measurements. Starting from 2 mm, the thickness of HZO film increases linearly from 0 to 50 nm. **b,** height variation comparison between the two HZO samples across the LSMO-HZO boundary (where the HZO film thickness starts). The height variations are computed from standard deviation of 2 $\mu$m height maps measured by SPM. **c-g,** height maps at four different sample locations corresponding to colored arrows in **a** for the sample-1. **h-l,** height maps at equivalent sample locations for the sample-2. The distances are calibrated by the first appearance of HZO particles between the two samples.

II. Results and discussions

**Thickness-Gradient Combinatorial Libraries**

The two Hf$_{0.5}$Zr$_{0.5}$O$_2$ (HZO) gradient libraries designed in this study provide an effective materials-engineering platform for probing the transition from ionic to ferroelectric electromechanical behavior within a single specimen. Each library consists of a uniform 20-nm-thick La$_x$Sr$_{1-x}$MnO$_3$ (LSMO) bottom electrode grown on SrTiO$_3$ (001) single-crystal substrates, followed by an HZO overlayer with a controlled thickness gradient spanning 0–50 nm. The La content $x$ was 0.8 for Sample-1 and 0.55 for Sample-2. To provide a non-ferroelectric reference region for subsequent scanning-probe measurements, the first 2 mm of the LSMO electrode was intentionally left uncovered during HZO deposition. The HZO thickness gradient was realized using continuous linear mask-translation movement during pulsed-laser deposition (PLD), with each iteration depositing a smooth gradient from zero to one full HZO unit cell (~4 Å). Deposition proceeded until the thickest edge reached the target thickness. Reflection high-energy electron diffraction (RHEED) was employed to monitor the surface structure of the films *in situ* to ensure the epitaxial growth, and the HZO film was found to have the orthorhombic structure. In order to study the effect of the bottom electrode on the properties of the gradient-thickness HZO, two different samples were fabricated with different LSMO resistivities. In particular, for Sample-1, La$_{0.8}$Sr$_{0.2}$MnO$_3$ was deposited at the substrate temperature of 760 °C, O$_2$ pressure of 100 mTorr (at a laser repetition rate of 1 Hz), which resulted in the relatively high film resistivity of $1.7 \times 10^{-2}$ $\Omega \cdot$ cm, while for the Sample-2, La$_{0.55}$Sr$_{0.45}$MnO$_3$ was deposited at the same substrate temperature and the laser repetition rate, but at 50 mTorr of O$_2$ pressure, which resulted in a lower resistivity of $1.7 \times 10^{-3}$ $\Omega \cdot$ cm. On both samples, the gradient-thickness HZO films were deposited under identical conditions as described above, ensuring that differences in electromechanical behavior could be evaluated in relation to the electrode fabrication conditions and transport properties.

| Sample # | LSMO composition | LSMO resistivity ($\Omega \cdot$ cm) | O$_2$ pressure (mTorr) | Substrate temperature (°C) | Laser repetition rate (Hz) |
|---|---|---|---|---|---|
| 1 | La$_{0.8}$Sr$_{0.2}$MnO$_3$ | $1.7 \times 10^{-2}$ | 100 | 760 | 1 |
| 2 | La$_{0.55}$Sr$_{0.45}$MnO$_3$ | $1.7 \times 10^{-3}$ | 50 | 760 | 1 |

**Table 1.** Growth conditions of the Sample-1 and Sample-2.

Figure 1a schematically illustrates the library design and the measurement geometry. The comparison of surface roughness extracted from SPM height maps (Figure 1b) confirms that Sample-2 exhibits smoother morphology than Sample-1 across the HZO–LSMO boundary. The sequence of topography images recorded at five representative positions (Figures 1c–g) reveals the difference in the mobility of HZO particles during growth for the two samples. For the Sample-1 (Figures 1c-g), there are structures forming even away from the LSMO-HZO boundary (where the HZO thickness wedge starts) on the LSMO side, probably due to the high mobility of HZO on the unoptimized LSMO layer. In contrast, the Sample-2 (Figures 1h-l) shows clean atomic terraces of LSMO away from the LSMO-HZO boundary, indicating the HZO particles are better adhered to the optimized LSMO and thus promote better nucleation. One notable observation is that on Sample-1, there are triangular terraces forming on the LSMO side at distance of 50-100 $\mu$m away from the first appearance of HZO particles (Figure 1e). In contrast, there are only flat and clean LSMO terraces at the same locations on the Sample-2. These 50-100 nm terraces are likely HZO (111) forming on top of the LSMO (001) film. The difference of the Sample-1 and Sample-2 in terms of the formation of HZO microstructures on the LSMO side could be explained by the difference in the mobility of HZO particles on the LSMO surface during growth, or by the different plasma plume and mask shadowing induced by the different $O_2$ pressure for the two samples (Table 1).

On the HZO-covered side, Sample-2 exhibits smaller but denser nucleated particles of HZO, reflecting higher sticking probability and more uniform nucleation that facilitate rapid coalescence into a continuous film. This dense and fine-grained morphology explains the smoother topography and reduced ferroionic behavior observed in the sample with LSMO with a lower resistivity. In addition, it's notable that the LSMO terraces start deforming, as indicated by the blurring in Figure 1k, even before the full coverage of HZO layer forms.

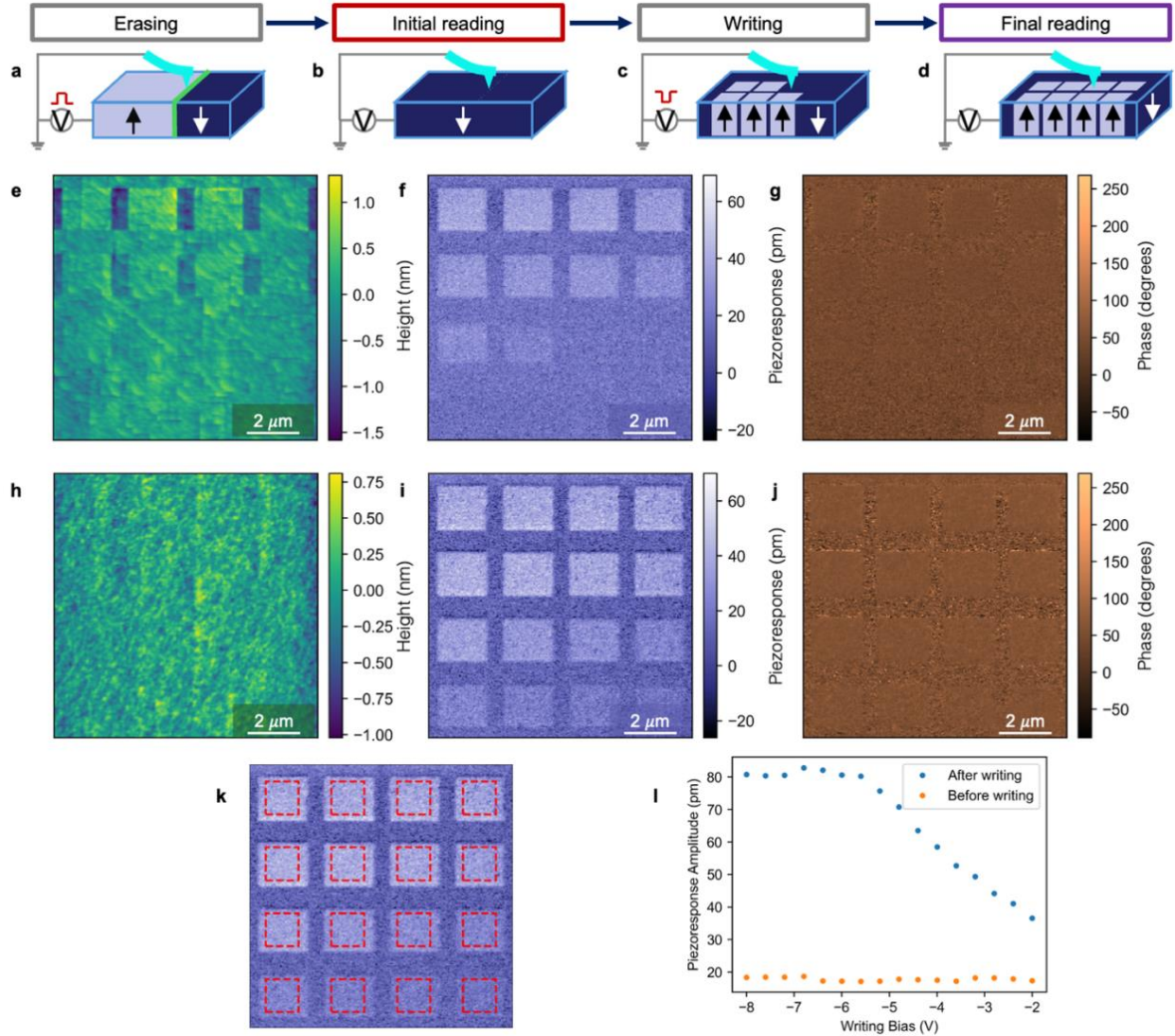

**Figure 2. Extract the topography and ferroelectricity switching induced by lithography on Sample-1. a-d,** schematic drawing showing the sequence of measurements at a single location: **a** scan with +8 V bias to erase natural domains, followed by **b** an initial reading with 0 V applied bias. **c** a series of smaller patterns are written from –2 V to –8 V and **d** the final written pattern is read out with 0 V again. **e-g,** example final reading scan of **e** height, **f** piezoresponse amplitude, and **g** piezoresponse phase on the LSMO side. **h-j,** example final reading scan on the HZO side (thickness = 11.25 nm). **k,** each red dashed box indicates the area of averaging for a specific writing voltage. **l,** the averaged piezoresponse amplitude can be plotted for both the initial reading and final reading scans to show the effect of lithography at different bias.

**Automated Scanning Probe Microscopy Protocol**

To interrogate nanoscale ferroelectric behavior systematically across these gradients, we implemented an automated PFM workflow[49] (Figure 2a–d). Each measurement cycle consisted of (a) a +8 V scan to erase the natural ferroelectric domains in the region, (b) an initial 0 V read of topography, piezoresponse amplitude, and piezoresponse phase, (c) a lithographic writing step using localized negative biases (–2 V → –8 V) in a predefined pattern, 8 x 8 in the 10 *μ*m region, and (d) a final 0 V read to capture the written state. This sequence was repeated at multiple positions along the thickness gradient, enabling spatially encoded, bias-resolved mapping of electrochemical and electromechanical hysteresis within a single automated run.

Figures 2e–j shows representative post-writing maps for height, piezoresponse amplitude, and piezoresponse phase both on the LSMO side (HZO thickness = 0) and within the HZO region (HZO thickness = 11.25 nm). The extracted average amplitude and phase values from each patterned square (Figure 2k–l) quantify the local switching response as a function of writing bias. This automation framework ensures identical tip trajectories and feedback parameters between experiments, minimizing operator bias and allowing statistically meaningful comparisons between samples and positions.

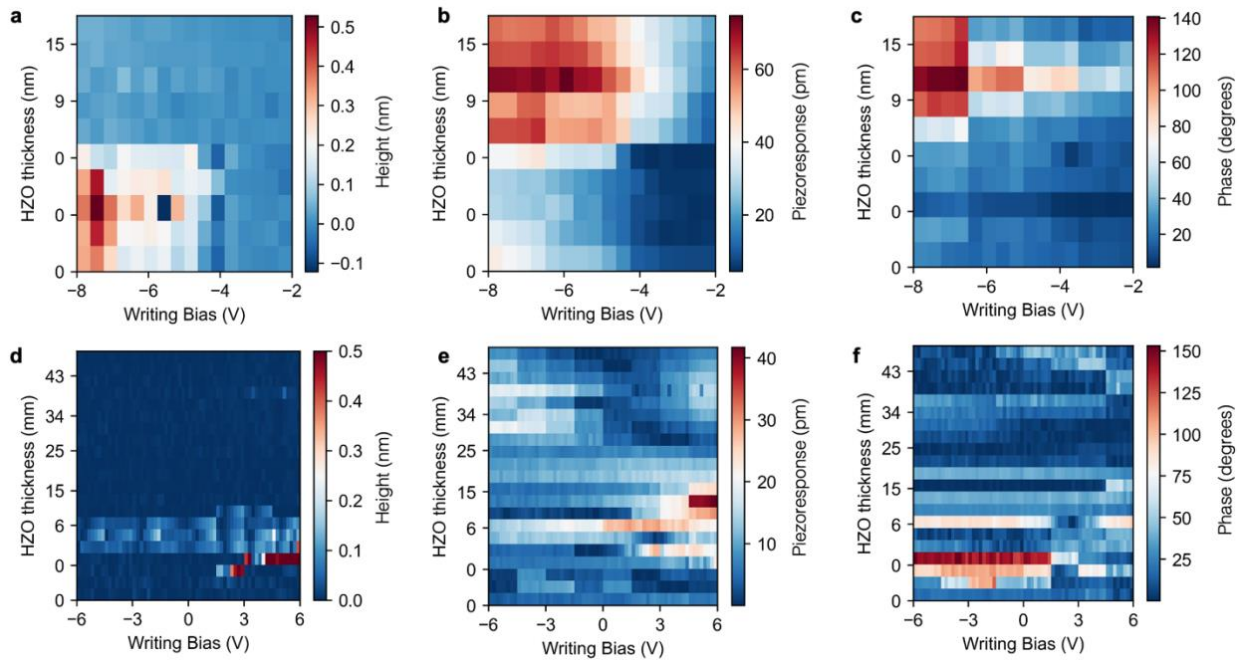

**Figure 3. Topography and piezoresponse switching at different thickness and bias. a-c,** change of **a** sample height, **b** piezoresponse amplitude, and **c** piezoresponse phase before and after the lithography scan across the Sample-1. At each location, the difference between final reading

and initial reading scans for each channel is computed as a horizontal line in the plot. **d-f,** similar plots for Sample-2.

**Ferroelectric versus Ionic Contributions**

For Sample-1, clear evidence of electrochemical activity was observed on the LSMO side. The biased lithography scans produced irreversible topographic deformation, indicating surface modification driven by ionic motion within the LSMO or by diffused HZO species reacting at the surface. These electrochemically induced height changes appeared only above a threshold bias ($|V| > 4$ V), consistent with field-activated ionic migration or electrochemical reaction rather than purely mechanical effects. On the HZO side of Sample-1, the irreversible topography changes were largely suppressed, and the piezoresponse phase maps revealed reversible contrast inversion characteristic of ferroelectric domain switching.

In contrast, Sample-2 exhibited a pronounced reduction in regions displaying irreversible electrochemical deformation, indicative of suppressed ferroionic activity. The scarcity of such features suggests that the topographic modifications observed in Sample-1 are more likely associated with mobile HZO adatoms or clusters and interfacial electrochemical processes rather than purely intrinsic lattice responses. Notably, the LSMO bottom electrode in Sample-2 has a lower resistivity than that of Sample-1. Because the two electrodes differ in both composition and growth oxygen pressure, this resistivity difference cannot be uniquely attributed to a single factor; however, it nevertheless represents a meaningful change in interfacial electrical boundary conditions. A more conductive electrode can enhance charge screening and reduce local electrochemical potential gradients, thereby limiting ionic migration and interfacial redox activity. Sample-2 shows reduced signatures of ferroionic behavior and a response dominated by reversible electromechanical contrast, consistent with intrinsic ferroelectric switching. These observations therefore suggest that electrode transport properties, regardless of their microscopic origin, play a key role in regulating the balance between ferroionic and ferroelectric mechanisms in gradient-thickness HZO heterostructures.

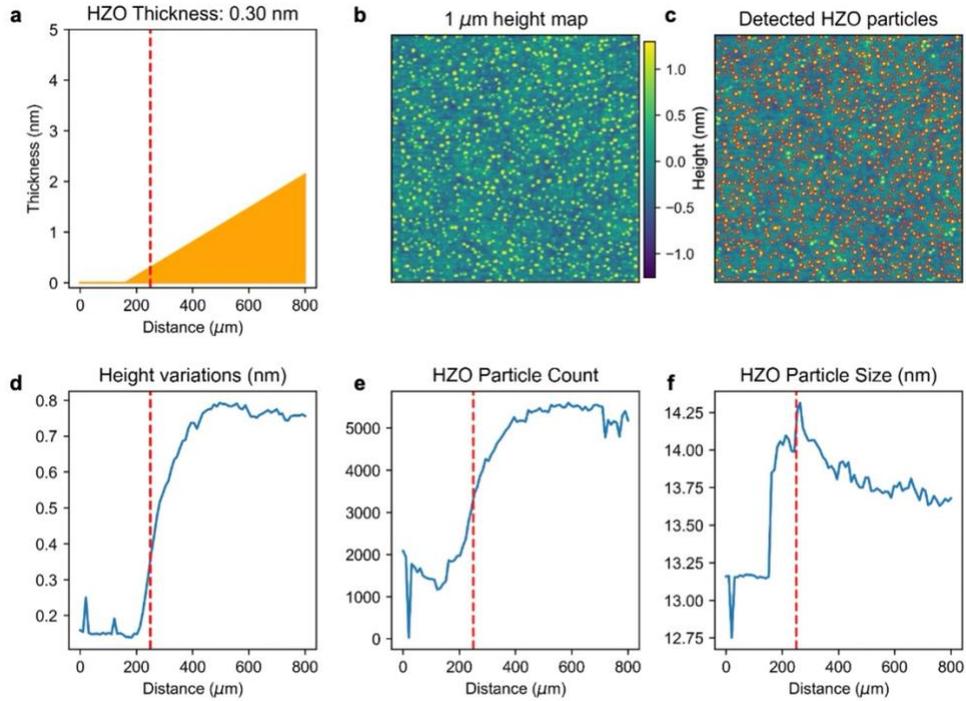

**Figure 4. Growth information extracted from topography height maps. a,** the thickness profile of HZO across the LSMO-HZO boundary in the Sample-2, with a step size of 10 $\mu$m. **b,** topography height map measured at the red dashed line in a. **c,** detected HZO particles are marked by red circles. A custom Log-G method is used to extract the locations, sizes, and number of particles presented in the height map. **d-f,** evolution of **d** sample roughness, **e** HZO particle numbers, and **f** detected HZO particle sizes across the LSMO-HZO boundary.

**Morphological Evolution Across the LSMO–HZO Boundary**

Figure 4 summarizes the quantitative analysis of HZO nucleation and surface evolution across the LSMO–HZO boundary in Sample-2. The local film thickness profile reconstructed from topography (Figure 4a) and the corresponding two-dimensional height map (Figure 4b) reveal a continuous transition from bare LSMO terraces to a fully coalesced HZO film. Individual HZO nuclei were identified and quantified using a Laplacian-of-Gaussian (Log-G) detection algorithm with geometric filtering (Figure 4c), allowing extraction of the spatial distributions of surface roughness, particle density, and particle size along the gradient.

The evolution of these metrics (Figures 4d–f) highlights several key trends that elucidate the early-stage growth mechanism. First, the measured surface roughness shows a strong positive correlation with the areal density of HZO particles, confirming that roughness can serve as an effective proxy for nucleation density and overall surface quality during HZO growth. Regions with sparse particle coverage correspond to atomically smooth terraces, while the onset of dense nucleation leads to a rapid increase in roughness until full surface coverage is achieved.

Second, the average particle size exhibits a distinct non-monotonic behavior: it increases sharply during the initial nucleation stage, reaching a maximum just before the LSMO surface becomes fully covered, and then gradually decreases as the particle count approaches saturation. This trend reflects a crossover from isolated island growth to coalescence and grain refinement. During early growth, high adatom mobility allows small nuclei to merge into larger aggregates, producing a transient size maximum. Once the surface becomes nearly continuous, further adatom arrival promotes infill between grains and smoothing of height variations, leading to a decrease in the apparent particle size and stabilization of surface roughness.

These observations collectively indicate that the HZO growth conditions in Sample-2 establish a balanced kinetic regime between adatom mobility and sticking probability—sufficient to promote dense nucleation without triggering uncontrolled island coarsening. This behavior is likely linked to the lower oxygen pressure used during deposition, which can modify plume energetics, surface diffusion lengths, and oxidation kinetics, thereby influencing nucleation density and island evolution. Consistent with this interpretation, the LSMO layer in Sample-2 exhibits a lower resistivity than that of Sample-1, a difference that may reflect variations in growth environment as well as composition. Nevertheless, the concurrent changes in oxygen pressure, transport properties, and surface morphology support the broader conclusion that deposition atmosphere plays a key role in regulating growth dynamics. The quantitative agreement between roughness and particle density further validates real-time topography metrics as rapid, non-destructive indicators of surface quality in growth-optimization workflows.

The strong coupling between surface morphology and nucleation dynamics has direct implications for the control of ferroelectric phase formation. Dense, fine-grained HZO films generated through uniform nucleation promote the stabilization of the orthorhombic ferroelectric phase via grain-confinement effects, while excessive coarsening or incomplete coverage favors non-ferroelectric monoclinic domains. The demonstrated correlation between roughness and particle statistics thus provides a quantitative pathway to incorporate surface metrics into autonomous feedback loops, enabling real-time optimization of film smoothness, grain size, and ultimately ferroelectric performance.

## Conclusion

Thickness-gradient combinatorial libraries coupled with automated PFM offer a robust strategy for deconvolving ionic and ferroelectric electromechanical signals within a single heterostructure. Across two HZO–LSMO libraries, we show that electrochemical deformation originates primarily from ionic activity at the exposed LSMO surface and optimized growth conditions suppress interfacial reactions by improving HZO adatom sticking and nucleation uniformity. In addition, quantitative topography metrics, like particle density, size evolution, and roughness, serve as reliable indicators of early growth dynamics and resulting ferroelectric functionality. These findings establish a direct link between nucleation physics, interfacial chemistry, and local switching behavior, enabling mechanism-resolved interpretation of nanoscale electromechanical maps.

Beyond this study, the demonstrated workflow provides a general framework for autonomous exploration of complex oxide libraries, real-time assessment of thin-film growth quality, and rapid screening of ferroelectric performance across multidimensional parameter spaces. Future directions include integrating real-time feedback into pulsed-laser deposition, expanding to ternary or dopant-gradient libraries, and coupling electromechanical mapping with spectroscopy and machine-learning–based structure–response models. Together, these advances will accelerate the discovery and optimization of ferroelectric and ferroionic materials for energy, memory, and microelectronic technologies.

## Acknowledgements

This work is supported by the center for 3D Ferroelectric Microelectronics Manufacturing (3DFeM2), an Energy Frontier Research Center funded by the U.S. Department of Energy (DOE), Office of Science, Basic Energy Sciences under Award Number DE-SC0021118.